\documentclass[twoside,twocolumn,9pt]{article}
\usepackage{extsizes}
\usepackage[super,sort&compress,comma]{natbib} 
\usepackage[version=3]{mhchem}
\usepackage[left=1.5cm, right=1.5cm, top=1.785cm, bottom=2.0cm]{geometry}
\usepackage{balance}
\usepackage{mathptmx}
\usepackage{sectsty}
\usepackage{graphicx} 
\usepackage{lastpage}
\usepackage[format=plain,justification=justified,singlelinecheck=false,font={stretch=1.125,small,sf},labelfont=bf,labelsep=space]{caption}
\usepackage{float}
\usepackage{fancyhdr}
\usepackage{fnpos}
\usepackage[english]{babel}
\addto{\captionsenglish}{%
  
}
\usepackage{array}
\usepackage{droidsans}
\usepackage{charter}
\usepackage[T1]{fontenc}
\usepackage[usenames,dvipsnames]{xcolor}
\usepackage{setspace}
\usepackage[compact]{titlesec}
\usepackage{hyperref}
\usepackage{amsmath,amssymb}

\usepackage{epstopdf}

\definecolor{cream}{RGB}{222,217,201}

\begin{document}

\pagestyle{fancy}
\thispagestyle{plain}
\fancypagestyle{plain}{
\renewcommand{\headrulewidth}{0pt}
}

\makeFNbottom
\makeatletter
\renewcommand\LARGE{\@setfontsize\LARGE{15pt}{17}}
\renewcommand\Large{\@setfontsize\Large{12pt}{14}}
\renewcommand\large{\@setfontsize\large{10pt}{12}}
\renewcommand\footnotesize{\@setfontsize\footnotesize{7pt}{10}}
\makeatother

\renewcommand{\thefootnote}{\fnsymbol{footnote}}
\renewcommand\footnoterule{\vspace*{1pt}%
\color{cream}\hrule width 3.5in height 0.4pt \color{black}\vspace*{5pt}} 
\setcounter{secnumdepth}{5}

\makeatletter 
\renewcommand\@biblabel[1]{#1}            
\renewcommand\@makefntext[1]%
{\noindent\makebox[0pt][r]{\@thefnmark\,}#1}
\makeatother 
\renewcommand{\figurename}{\small{Fig.}~}
\sectionfont{\sffamily\Large}
\subsectionfont{\normalsize}
\subsubsectionfont{\bf}
\setstretch{1.125} 
\setlength{\skip\footins}{0.8cm}
\setlength{\footnotesep}{0.25cm}
\setlength{\jot}{10pt}
\titlespacing*{\section}{0pt}{4pt}{4pt}
\titlespacing*{\subsection}{0pt}{15pt}{1pt}

\fancyfoot{}
\fancyfoot[LO,RE]{\vspace{-7.1pt}\includegraphics[height=9pt]{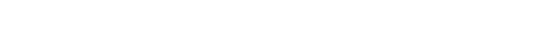}}
\fancyfoot[CO]{\vspace{-7.1pt}\hspace{13.2cm}\includegraphics{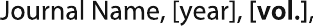}}
\fancyfoot[CE]{\vspace{-7.2pt}\hspace{-14.2cm}\includegraphics{head_foot/RF}}
\fancyfoot[RO]{\footnotesize{\sffamily{1--\pageref{LastPage} ~\textbar  \hspace{2pt}\thepage}}}
\fancyfoot[LE]{\footnotesize{\sffamily{\thepage~\textbar\hspace{3.45cm} 1--\pageref{LastPage}}}}
\fancyhead{}
\renewcommand{\headrulewidth}{0pt} 
\renewcommand{\footrulewidth}{0pt}
\setlength{\arrayrulewidth}{1pt}
\setlength{\columnsep}{6.5mm}
\setlength\bibsep{1pt}

\makeatletter 
\newlength{\figrulesep} 
\setlength{\figrulesep}{0.5\textfloatsep} 

\newcommand{\topfigrule}{\vspace*{-1pt}%
\noindent{\color{cream}\rule[-\figrulesep]{\columnwidth}{1.5pt}} }

\newcommand{\botfigrule}{\vspace*{-2pt}%
\noindent{\color{cream}\rule[\figrulesep]{\columnwidth}{1.5pt}} }

\newcommand{\dblfigrule}{\vspace*{-1pt}%
\noindent{\color{cream}\rule[-\figrulesep]{\textwidth}{1.5pt}} }

\makeatother

\twocolumn[
  \begin{@twocolumnfalse}
{\includegraphics[height=30pt]{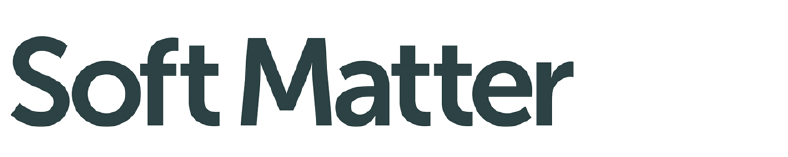}\hfill\raisebox{0pt}[0pt][0pt]{\includegraphics[height=55pt]{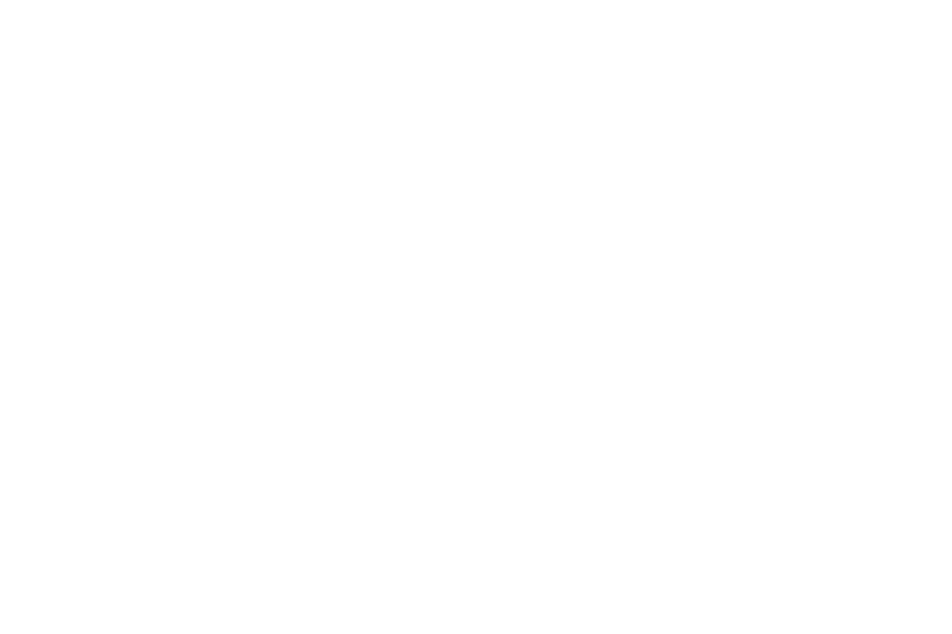}}\\[1ex]
\includegraphics[width=18.5cm]{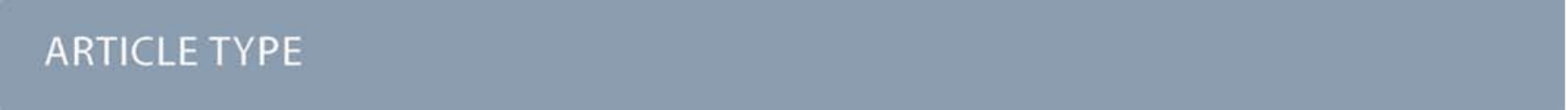}}\par
\vspace{1em}
\sffamily
\begin{tabular}{m{4.5cm} p{13.5cm} }

\includegraphics{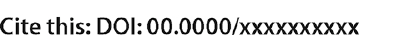} & \noindent\LARGE{\textbf{Microstructure {of the near-wall layer} of filtration-induced colloidal assembly}} \\
\vspace{0.3cm} & \vspace{0.3cm} \\

 & \noindent\large{Mohand Larbi Mokrane,\textit{$^{a,b}$ $^{\dagger}$} T\'erence Desclaux,\textit{$^{a}$ $^{\dagger}$}  Jeffrey Morris,\textit{$^{c}$} Pierre Joseph\textit{$^{b}$} and Olivier Liot\textit{$^{a,b}$ $^{\ast}$}} \\

\includegraphics{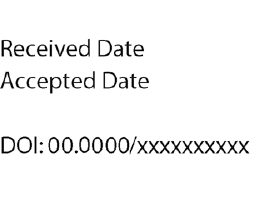} & \noindent\normalsize{This paper describes an experimental study of filtration of a colloidal suspension using microfluidic devices. A suspension of micrometer-scale colloids flows through parallel slit-shaped pores at fixed pressure drop. Clogs and cakes are systematically observed at pore entrance, for variable applied pressure drop and ionic strength. Based on image analysis of the layer of colloids close to the device wall, global and local studies are performed to analyse in detail the {near-wall layer} microstructure. Whereas global porosity {of this layer} does not seem to be affected by ionic strength and applied pressure drop, a local study shows some heterogeneity: clogs are more porous at the vicinity of the pore than far away. An analysis of medium-range order using radial distribution function shows a slightly more organized state at high ionic strength. This is confirmed by a local analysis using two-dimension continuous wavelet decomposition: the typical size of crystals of colloids is larger for low ionic strength, and it increases with distance from the pores. We bring these results together in a phase diagram involving colloid-colloid repulsive interactions and fluid velocity.} \\


\end{tabular}

 \end{@twocolumnfalse} \vspace{0.6cm}

  ]

\renewcommand*\rmdefault{bch}\normalfont\upshape
\rmfamily
\section*{}
\vspace{-1cm}


\footnotetext{\textit{$^{a}$~Institut de Mécanique des Fluides de Toulouse, Université de Toulouse, CNRS, Toulouse, France.}}
\footnotetext{\textit{$^{b}$~LAAS-CNRS, Université de Toulouse, CNRS, Toulouse, France.}}
\footnotetext{\textit{$^{c}$~Levich Institute and Chemical Engineering, CUNY
City College of New York, New York, USA.}}
\footnotetext{\textit{$^\dagger$~Co-first: both authors contributed equally to this work.}}
\footnotetext{\textit{$^\ast$~Corresponding author: olivier.liot@imft.fr.}}





\section{Introduction}

Filtration is a central process in many fields of application, such as water treatment or bioprocessing. Membrane fouling and the induced hydraulic resistance enhancement are key problems for improving filtration processes (power consumption, membranes' lifetime). When a suspension flows through a membrane, many objects (colloids, cells, bacteria, aggregates) can accumulate at the membrane surface. The efficiency of the filtration decreases as the hydraulic resistance induced by accumulated particles rises. This has a dramatic impact on oil recovery\cite{tavakkoli2015}, ink-jet printing\cite{fuller2002}, biodetection\cite{iv2015} or water infiltration in soils\cite{zhang2012} and even in brain diseases\cite{bonazzi2018}. For decades, filtration studies have focused on membrane scale and global hydraulic response\cite{in-soung2002,guo2012} or membrane materials\cite{meng2009}. Since the advent of microfluidics at the beginning of the 21\textsuperscript{st} century, the study of microfiltration has become a very active field of research\cite{vollebregt2010}. The use of microfluidic devices is indeed an efficient way to understand the key microscopic mechanisms involved during the filtration process. Wyss \emph{et al.}\cite{wyss2006} were precursors in this domain, followed by many others, as detailed in a recent review\cite{dressaire2017}. 

Different strategies and scales have been explored. At particle level, the first steps of clogging process have been studied using confocal microscopy \cite{dersoir2017,dersoir2019}. Adhesion of one particle on a pore surface involves different parameters or mechanisms such as pore geometry\cite{duru2015,laar2016} or Brownian motion\cite{cejas2017}. At pore scale, many studies proposed experimental and/or theoretical-numerical approaches\cite{wyss2006,bacchin2011,agbangla2014,dersoir2015,robert_de_saint_vincent2016,sauret2018} to clogging dynamics. The upscaling of pore-scale results towards membrane scale is a recent development using parallel pore arrays\cite{liot2018-1,van_zwieten2018,sauret2018} or more complex porous-like media\cite{gerber2018,gerber2019}. This revealed some complex mechanisms of ``cross-talk'' between pores during filtration processes. Especially, previous work of our team\cite{liot2018-1} revealed that pore redistribution by Brownian diffusion accelerates the clogging process. The proposed model is based on assumptions about clog hydraulic permeability. 

In such (sub-)micrometric devices, direct measurements of flow rate under given pressure drop are not accessible using commercial apparatus. Some indirect measurements are possible\cite{coyte2017}. But a direct analysis of the clog micro-structure can provide insights on clog permeability. Up to now, the first studies of pore microstructure have been performed for soft particle microfiltration such as microgel\cite{linkhorst2016,bouhid_de_aguiar2017,bouhid_de_aguiar2018,linkhorst2019}. These studies focused on microgels deformation, but they observed different structures of the filtration cake with amorphous or crystalline behaviours depending on forcing. To our knowledge, there is no systematic study of colloidal clog microstructure. 

Contrary to deformable microgels, colloidal particles of the form we study are essentially rigid and the physico-chemical surface properties are crucial to understand the way they accumulate under external forcing. At colloidal scale ($\lesssim 1\,\mu$m), typical interaction scales are similar to particle diameter. {In addition, Brownian diffusion brings an extra energy source which can allow energy barrier crossing necessary for adhesion; or influence particle spatial organisation\cite{vicsek1984}. }

{The Derjaguin, Landau, Verwey and Overbeek (DLVO) theory, consisting of a superposition of van der Waals interactions and electrostatics in an aqueous solution, is the standard approach to explain colloid-colloid or colloid-wall interactions\cite{hunter2013,van_de_ven1989,israelachvili1991,schoch2008}.}









{The repulsive colloid-colloid or surface-colloid interactions are caused by Electrical Double Layer (EDL) interactions\cite{van_de_ven1989,israelachvili1991}. The zeta potential $\zeta$ corresponds to the electric potential at the slip plane between the two sub-layers composing the EDL. The typical thickness of the total layer, the Debye length, corresponds to electrostatic surface charge screening. It is expressed as: }

\begin{equation}
    \kappa^{-1}=\sqrt{\dfrac{\epsilon_r\epsilon_0k_BT}{e^2\sum_i\rho_{\infty,i}z_i^2}},
\end{equation}

\noindent where $\epsilon_r$ is the relative dielectric constant of the fluid, $\epsilon_0$ the vacuum electric permittivity, $k_B$ the Boltzmann constant, $T$ the temperature, $e$ the elementary charge, $\rho_{\infty,i}$ the bulk concentration of ion $i$ and $z_i$ its valence. Consequently, the Debye layer is inversely proportional to bulk concentration to the power $1/2$: the more concentrated the solution, the thinner the EDL. The ionic strength is defined as:

\begin{equation}
    I=\dfrac{1}{2}\sum_i\rho_{\infty,i}z_i^2.
\end{equation}

\noindent So we can write $\kappa^{-1}\propto 1/\sqrt{I}$.








{The total interaction energy is the combination of electrostatic and van der Waals interactions.  When the distance between two objects is reduced, a secondary minimum appears {(which is known to have a significant impact on filtration\cite{tufenkji2005})}. The energy profile contains a primary minimum at very low inter-particle distance. A particle in this primary minimum is at a much lower energy than that needed to escape, thus it is adhered through van der Waals interactions.}




The second main mechanism at play in assembling colloids is Brownian motion.  When a colloid is advected by surrounding flow, the competition between advection and Brownian diffusion can be interpreted using the dimensionless P\'eclet number: 

\begin{equation}
    Pe=\dfrac{\text{advective transport}}{\text{diffusive transport}}=\dfrac{3\pi\eta d^2U}{k_BT},
\end{equation}

\noindent with $\eta$ the fluid viscosity, $d$ the particle diameter and $U$ the typical advection velocity. A Péclet number smaller than 1 means that diffusion is more important than advection, which can easily be reached when colloids are advected at a decreasing velocity during membrane fouling process. {This mechanism (shear, Brownian motion) was added to the DLVO theory\cite{zaccone2009,zaccone2010}, as well as hydrogen bonds\cite{van_oss1986}, to build an extended DLVO theory (XDLVO).}


This theoretical framework is essential to study microfiltration of colloids.  In this work we use an experimental approach to investigate the microstructure of filtration-induced colloidal assembly in model microfluidic systems. We flow a suspension which progressively clogs the pore. Spinning-disk confocal microscopy is used to determine individual positions of accumulating particles during clogging, for different ionic strengths and applied pressure drops. We propose an analysis of the global clog properties before focusing on more local tools to understand the heterogeneity in the spatial organisation of the colloidal particles. 

\section{Experimental methods}

We use a model-system approach in order to tightly control the different parameters such as pore dimension, colloid size and properties or ions concentration. 

\subsection{Particle suspension}

Our experiments consist in flowing a suspension through microfabricated slits to observe accumulation of colloids at their entrance. We use polystyrene beads (diameter $d=1\pm0.02\,\mu$m, density 1.05\,g.cm$^{-3}$) volume-loaded with fluorophore, so they can be easily observed using fluorescence microscopy. {Very nearly monodisperse particles used insures that the particle diameter distribution should not influence on clog structure.} The particle surfaces are carboxylate-modified leading to a negative surface charge, with measured zeta-potential near $-50\,$mV ($pH\approx6$); measurements performed by dynamic light scattering on a \emph{Malvern} ``Zeta-Sizer'' instrument. 

We use suspensions with volume fraction of $\phi=4\times10^{-5}$. For the 1\,$\mu$m-diameter particles we use, this corresponds to $7.9\times10^4$ particles per $\mu$L. By adding potassium chloride, two ionic strengths are used to compare repulsive interaction effects: $I=0.5\,$mM and $I=5\,$mM. This corresponds to $\kappa^{-1}=13\,$nm and $\kappa^{-1}=4\,$nm respectively. {Note that for the carboxylate-modified latex particles we use, the Critical Coagulation Concentration in presence of ions K$^+$ is $51$\,mM\cite{oncsik2014}. Thus we work in a salt concentration range far from spontaneous particle coagulation.}  The measured pH of the suspension is about 6.

{In the ionic strength range we used, the force profile between two carboxylate-modified latex particles versus interparticle distance reveals an electrostatic barrier whose height is about 0.5--1\,nN, and without secondary minimum\cite{elzbieciak-wodka2014}. If we consider the glass-particle interaction, the profile is very similar with a barrier of the same order of magnitude\cite{assemi2006}.}


\subsection{Microfluidic device}

\begin{figure}[!ht]
\centering
\includegraphics[width=0.9\linewidth]{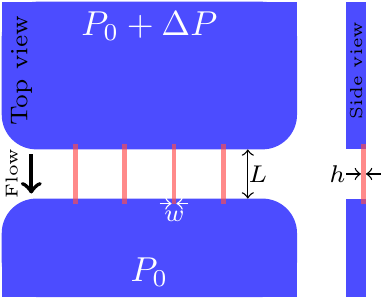}\\

\vspace{0.5cm}

\includegraphics[width=1\linewidth]{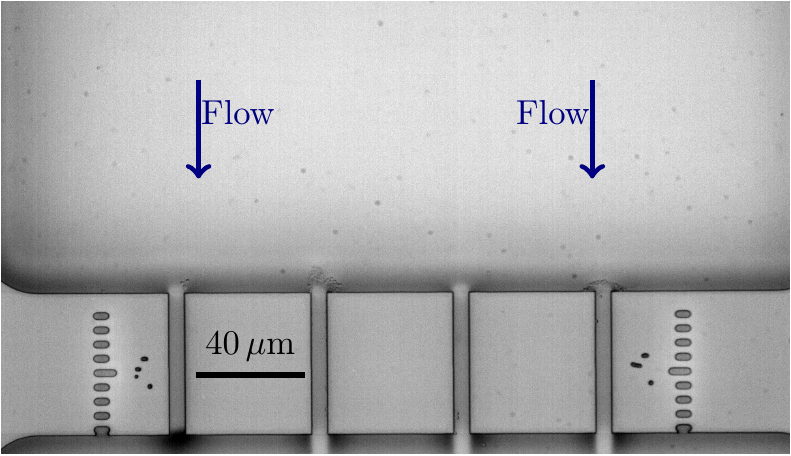}
\caption{Top: sketch of the microfluidic chip, slits (red) connect the microfluidic channel (blue); slits are connected to microchannel corners. Top view is on the left, side view is on the right. Bottom: micrograph of the slits, 40$\times$ magnification.}
\label{fig:design_image_blanche}
\end{figure}

A two-level microfabricated device is used to mimic membrane constrictions. Four parallel slits of cross-section $w\times h=5\times3.2\,\mu \text{m}^2$ and length $L=50\,\mu$m are formed using plasma etching in silicon. Separated by 50\,$\mu$m (center to center), they are supplied using inlet and outlet microfluidic channels (depth $18\,\mu$m, width $300\,\mu$m). A 170$\,\mu$m-thick borosilicate glass plate covers the silicon-etched wafer, sealed using anodic bonding. In our pH and saline conditions, the zeta potential of microfluidic chip materials is about -30\,mV for silicon\cite{bousse1991} and -55\,mV for borosilicate\cite{erickson2000}. {In the range of ionic strength we consider, surface charge of both silicon and borosilicate are not fully screened\cite{oh1999,erickson2000}.} Consequently, the EDL interactions between surfaces and beads will be repulsive. The design is presented in figure \ref{fig:design_image_blanche}\,(top). Due to the manufacturing process, the slits connect the corners of the cross-section of the inlet and outlet microchannels. Figure \ref{fig:design_image_blanche}\,(bottom) shows a bright-field micrograph of the microfluidic device we used. At 40$\times$ magnification the slits can easily be observed. 



The flow is imposed using a \emph{Fluigent} pressure controller in the range $\Delta P=$10--100\,mbar with precision under 0.1\,mbar. In the absence of any clog, this corresponds to a flow rate of $Q=$37--370\,nL.min$^{-1}$. {Note that the hydraulic radius $r_h$ of one slit and of the microchannel are related as: }

\begin{equation}
    {r_h^{microchannel}=8.7\,r_h^{slit}}
    \end{equation}

{Thus, as the hydraulic resistance scales as $r_h^{-4}$, we consider that the applied pressure drop is entirely a result of flow through the slits.}

\subsection{Clog observation}

\begin{figure}[!ht]
\centering
\includegraphics[width=0.9\linewidth]{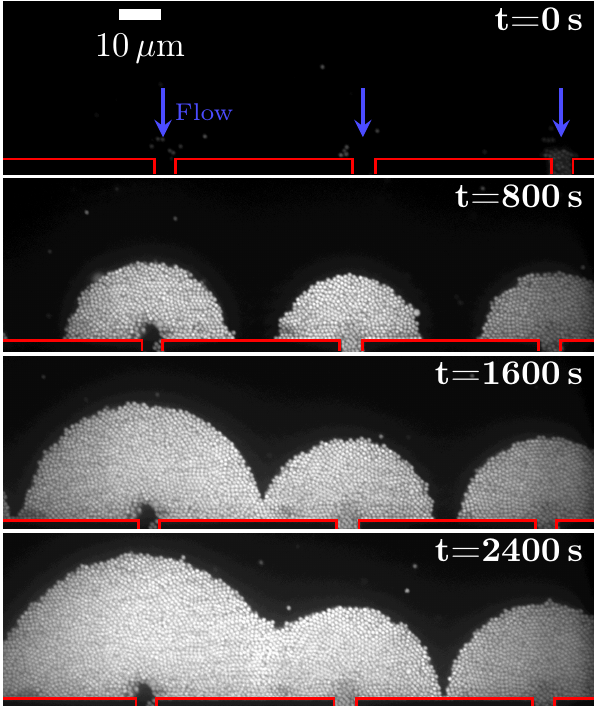}
\caption{Confocal micrography timelapse of the clog formation. Red lines represent the slits limits. $\Delta P=20\,$mbar and $I=5\,$mM. The small clog visible at the right-hand pore entrance ($t=0\,$s) is due to particles flowing in the device during experiment installation. It will not affect the clog microstructure, only the clogging trigger.}
\label{fig:image_clog}
\end{figure}

The clogging dynamics and clog structure are observed using confocal fluorescence microscopy (63$\times$ magnification). Unfortunately, because of non-transparency of the beads, we cannot access the 3D structure of the clogs. {Confocal microscopy enables removal of background light originating from out-of-plane particles. Thus, the particles at the wall layer are well resolved (see figure \ref{fig:image_clog}), permitting accurate particle detection.}

\begin{figure}[!ht]
\centering
\includegraphics[width=0.9\linewidth]{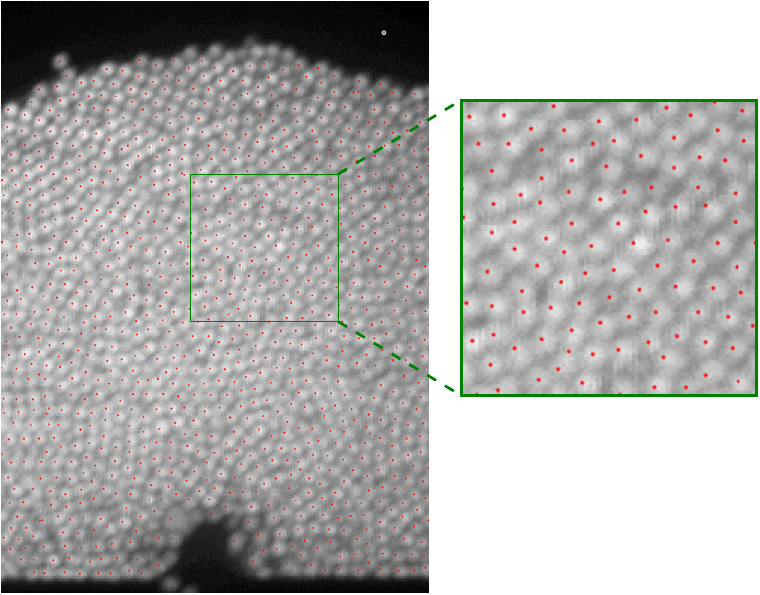}
\caption{Detection of the clog particles, with zoom on a portion of the clog. Red points correspond to detected particles' position.}
\label{fig:detection_particules}
\end{figure}

Clog formation is recorded with a frequency of 2 frames per minute.  Figure \ref{fig:image_clog} shows a sequence of images at increasing time under flow of the clog formation, recorded by confocal microscopy. Clogging is stopped after 2400\,s for each experiment, so that a similar amount of suspension flows through the device during the experiment. {The clog growth dynamics is beyond the scope of this article; some insights into this topic can be found in our previous work\cite{liot2018-1}. The ``final'' size of a clog will depend on the (stochastic) first steps of the clogging process. Consequently, when the experiment is stopped, some clogs have reached their maximal size whereas other are still growing.} We use custom \emph{Python} scripts to analyze the images and allow detection of both clog contour and particle position with sub-pixel precision. Figure \ref{fig:detection_particules} shows an example of particle detection within a clog.

{Due to the inability to unclog microfluidic devices in a satisfactory way, each experiment was made in a new device, adding complexity to the experimental process. A total of 11 experiments are usable, at different pressure drops and two ionic strengths. Three to four clogs are visible on each experiment, leading to about 40 different clogs. Each point and curve presented below is an average of three or four clogs extracted from the same experiment.}

\section{Results}


Since we can only access the wall layer -- the one in contact with the borosilicate glass plate -- {a bias on spatial organisation could be induced by this wall}. {However the similar zeta potential of colloids and wall should limit this bias, if it exists, to geometric and confinement factors.} Moreover, the trends shown here are generic to filtration processes, because they are related to particle/particle interactions, of the same nature as the ones with the walls of our devices. Observations detailed below remain related to physical and physico-chemical phenomena at play in filtration. Two main parameters are discussed: applied pressure drop and ionic strength.



\subsection{Global {wall layer} properties}

We start our analysis with global properties of the {wall layer of the} clog obtained by microfiltration. Two attributes can be computed: apparent porosity and radial distribution function.

We define the clog -- at least the wall layer -- apparent porosity as: 

\begin{equation}
    \epsilon=1-\dfrac{V_{particles}}{V_{clog}}.
\end{equation}

\noindent $V_{particles}$ is the total volume of particles included in the wall layer of the clog. If we note $N$ the number of particles, we get $V_{particles}=N\pi d^3/6$. The wall layer clog volume $V_{clog}$ is estimated as the volume defined by the clog area $A_{clog}$ on a thickness $d$: $V_{clog}=dA_{clog}$. Finally we have: 

\begin{equation}
    \epsilon=1-N\dfrac{\pi d^2}{6\,A_{clog}}.
\end{equation}

\begin{figure}[!ht]
\centering
\includegraphics[width=0.9\linewidth]{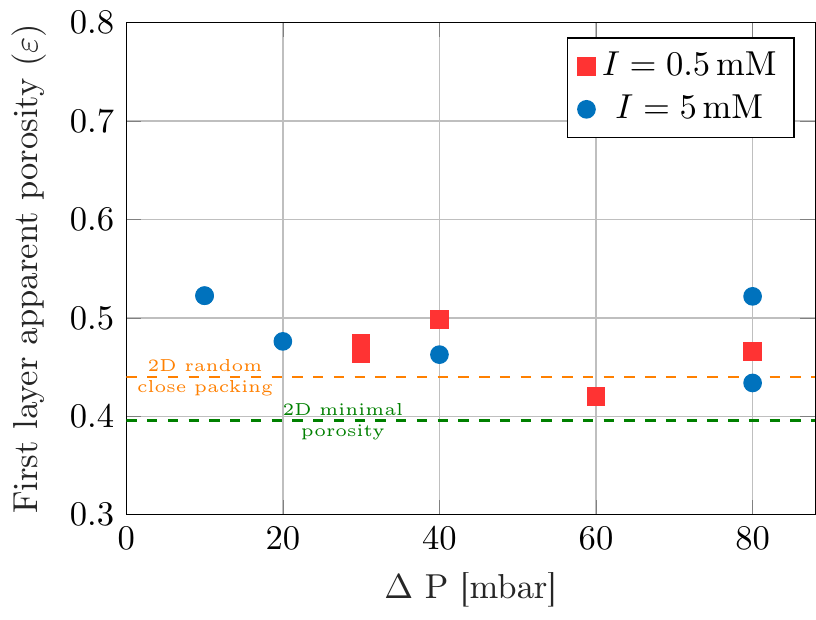}
\caption{Apparent porosity of the wall layer of the clog, as a function of the applied pressure drop and for two different ionic strengths. The two dashed lines represent the random close packing and minimal porosity for a 2D assembly of monodisperse spheres.}
\label{fig:porosity_global}
\end{figure}

\noindent Both $N$ and $A_{clog}$ are computed directly from the clog pictures (figures \ref{fig:image_clog} and \ref{fig:detection_particules}). Figure \ref{fig:porosity_global} shows the apparent porosity for different applied pressure drops and two ionic strengths (0.5\,mM and 5\,mM). We do not observe a visible trend when changing $I$ and applied pressure drop which means that the mean porosity of the wall layer is not affected by these two parameters and remains in the range 0.42--0.52. It is a bit higher than the expected porosity for a 2D random packing of monodisperse spheres. Intuitively, applying relatively weak pressure drop allows particles to find a better location, so it could lead to a less porous clog, which is not the case for these results. Actually, when we study the local porosity (see next section), some spatial heterogeneities appear.



To go further than the ratio between void and solids in the clog, we analyse the spatial organisation of the particles in the wall layer of the clog. We consider the Radial Distribution Function (RDF) which gives insights about average order at a distance $r$ from any particle. It can be expressed as (see e.g. Saw \emph{et al.}\cite{saw2012}):

\begin{equation}
    g(r)=\sum_{i=1}^N\dfrac{\psi_i(r)/N}{(N-1)\left(\dfrac{dS_r}{A_{clog}}\right)},
\end{equation}

\noindent where $\psi_i(r)$ is the number of particle centers included in a shell between $r-\delta r$ and $r+\delta r$. $N$ is the the total number of particles in the clog of area $A_{clog}$. $dS_r$ represents the shell surface. Edge effects represent the main limitation of this formula. Our clogs have indeed specific and non-regular shapes. To bypass this problem, we adapted a method proposed by Larsen \& Shaw\cite{larsen2018}.

\begin{figure}[!ht]
\centering
\includegraphics[width=0.9\linewidth]{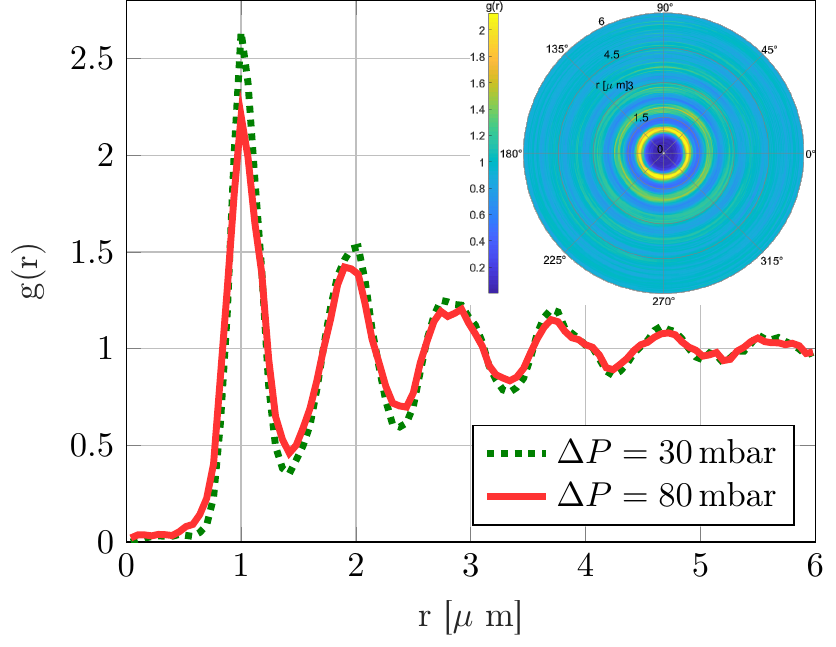}
\caption{Radial distribution function for two applied pressure drops at $I=0.5\,$mM. {Inset: RDF conditioned by direction angle (see text for details).}}
\label{fig:RDF_0.5}
\end{figure}

~~

\begin{figure}[!ht]
\centering
\includegraphics[width=0.9\linewidth]{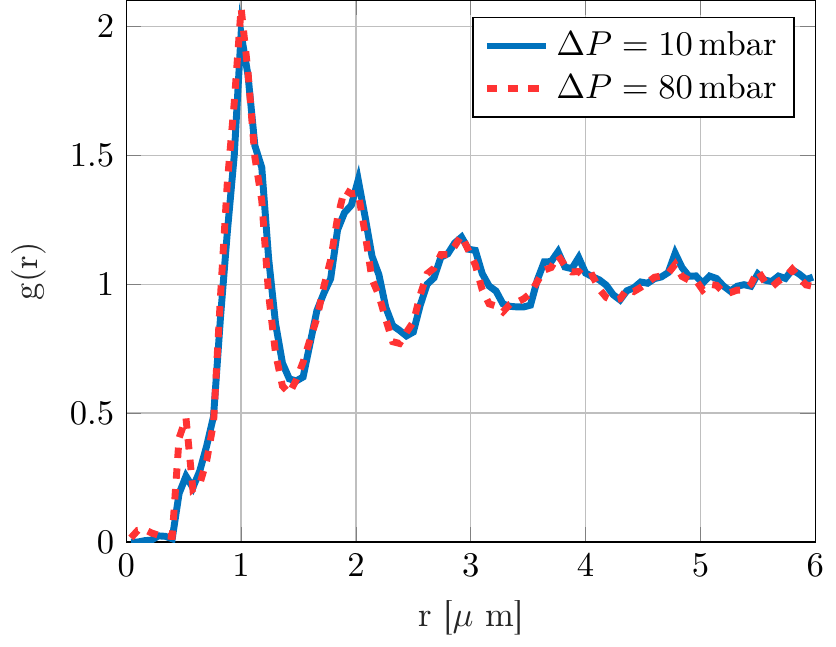}
\caption{Radial distribution function for two applied pressure drops at $I=5\,$mM. The blip observed at $r=0.5\,\mu$m is probably related to aliasing.}
\label{fig:RDF_5}
\end{figure}

We computed the RDF for different configurations $(I,\Delta P)$ on the whole cake{'s wall layer} formed at the pore entrance. Figures \ref{fig:RDF_0.5} and \ref{fig:RDF_5} show the RDF for $(I=0.5\,\text{mM},~\Delta P=[30;80]\,\text{mbar})$ and $(I=5\,\text{mM},~\Delta P=[10;80]\,\text{mbar})$ respectively. {Inset in figure \ref{fig:RDF_0.5} shows the RDF ($\Delta P=80\,$mbar, $I=0.5\,$mM) computed for different orientations. The angular domain is divided in 9 segments and nine RDF are computed by selecting only particles in each angular segment. This leads to nine RDF. Results are angle-interpolated to obtain a radial chart where color represents the $g(r)$ amplitude. This shows an isotropy of colloids spatial organisation.}

\begin{figure}[h!]
\centering
\includegraphics[width=0.98\linewidth]{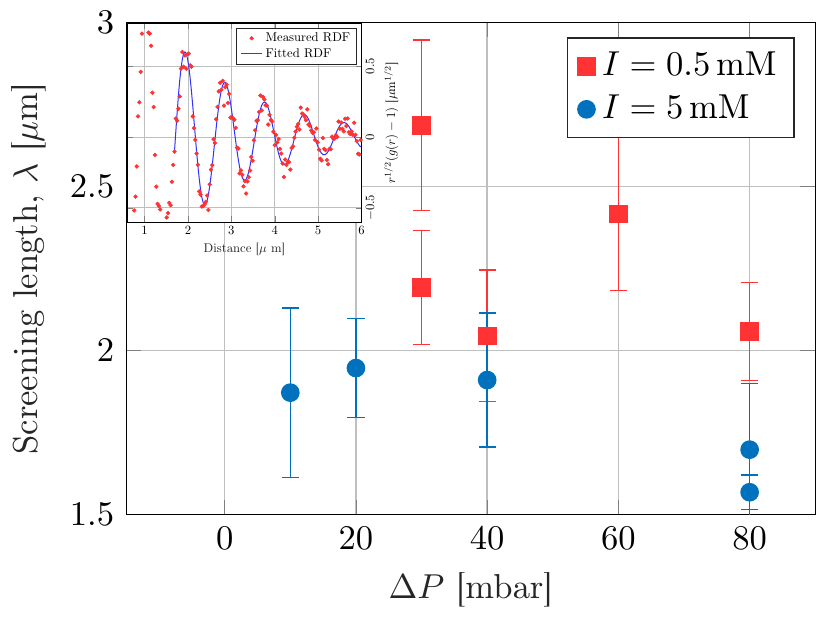}
\caption{Screening length as a function of applied pressure drop for the two different ionic strengths. Error bars correspond to fit's 95\%-confidence bounds. Inset: normalized RDF for $\Delta P=80$\,mbar and $I=0.5\,$mM. The solid line represents the fit obtained with equation \ref{eq:RDF_th}.}
\label{fig:RDF_fit}
\end{figure}

{For the total RDF, }we do not observe a clear difference when changing the applied pressure drop at fixed ionic strength. The comparison of $I=0.5\,$mM and $I=5\,$mM (not shown here) at given pressure drop does not reveal notable difference either. The small differences in peak positions and heights, visible in figure \ref{fig:RDF_0.5}, are difficult to analyse because of RDF resolution. Nevertheless, these RDF can give some insights about average spatial organisation of particles in a filtration clog/cake. The oscillations observed on the RDF of a 2D amorphous crystal can be modeled as\cite{outhwaite1975,bodapati2006,levashov2005}: 

\begin{equation}
g(r)-1=\dfrac{K}{r^{1/2}}\exp\left(-\dfrac{r}{\lambda}\right)\sin\left(\dfrac{2\pi r}{D}+\phi\right),
    \label{eq:RDF_th}
\end{equation}

\noindent where $K$, $\lambda$, $D$ and $\phi$ are constant coefficients. $D$ represents the oscillation period while $\lambda$ is a ``screening length'' quantifying the decay of a local spatial organisation (not to be confused with the Debye screening length). The higher $\lambda$, the more spatially organized the material. This expression is an asymptotic behaviour and is a signature of the medium-range order of the considered 2D assembly. For this reason, we fit the experimental RDF excluding the first peak. Inset in figure \ref{fig:RDF_fit} shows an example of fitted RDF (actually of $r^{1/2}(g(r)-1)$).

Figure \ref{fig:RDF_fit} presents the screening length $\lambda$ versus the applied pressure drop $\Delta P$ for $I=0.5$\,mM and $I=5\,$mM. Two results emerge from this plot. First, the screening length is higher for low ionic strength. Second, screening length seems to decrease at high pressure drop. This could be the signature of the competition between hydrodynamic forcing and electrostatic repulsion. At low $I$ and $\Delta P$, particle adhesion will be prevented or delayed allowing for colloids to self-organize, whereas at high $I$ and $\Delta P$, adhesion is facilitated which leads to a more amorphous clog. This interpretation will be detailed in the section \ref{sec:discussion}. In fact our results averaged on the whole clog{'s wall layer} are rather dispersed. This is due to heterogeneity of the clogs: very amorphous regions coexist with perfectly crystalline ones. {This was already observed in filtration cakes during microgel filtration\cite{linkhorst2016}. However our situation is quite different because we study hard particles, and physico-chemical colloid-colloid interactions differ widely from microgel-microgel ones.}


\begin{figure*}[h!]
\centering
\includegraphics[width=0.9\linewidth]{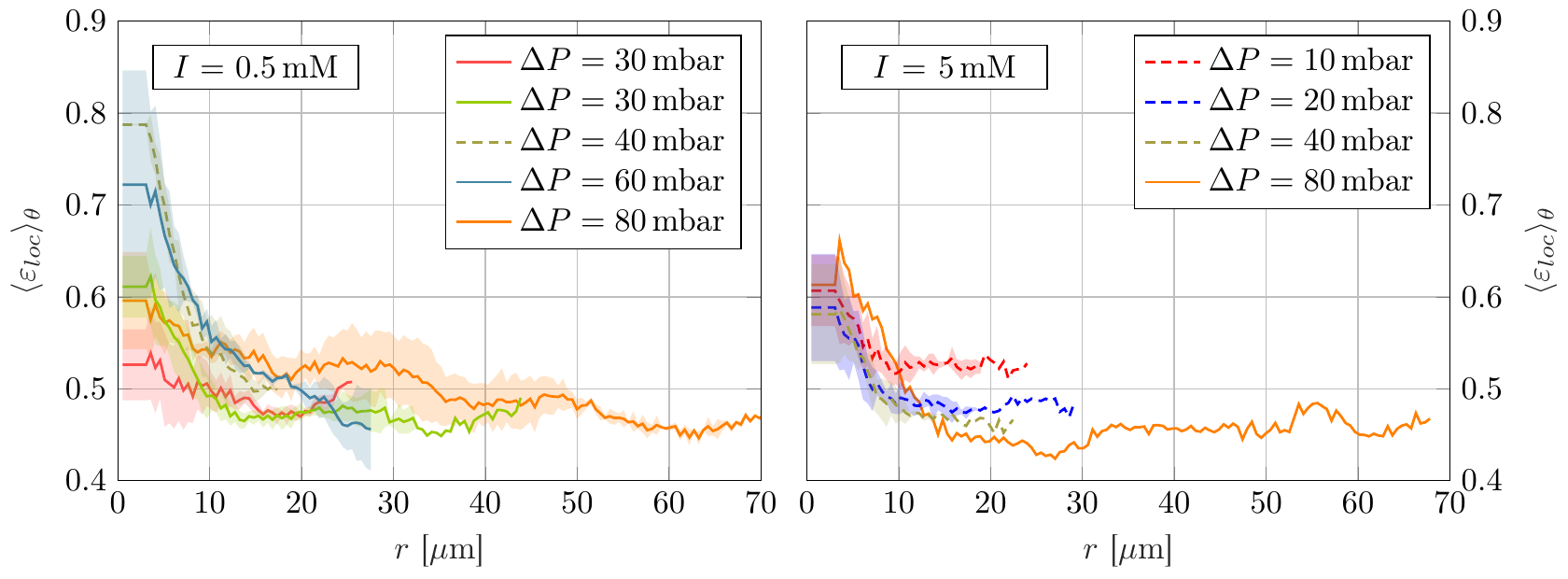}
\caption{Angularly averaged local porosity for (left) $I=0.5\,$mM and (right) $I=5\,$mM  as a function of the distance from the pore entrance. The uncertainty range corresponds to standard deviation obtained with several clogs in the same experiment (when possible). Short plateau for small $r$ is an extrapolation of the smallest exploitable radius $r$.}
\label{fig:porosity_average}
\end{figure*}
\subsection{Local analysis}

In order to quantify these heterogeneities, we propose a local analysis of clog microstructure, considering the local porosity and local colloid organisation. 

\subsubsection{Clog local porosity}

\begin{figure}[h!]
\centering
\includegraphics[width=0.9\linewidth]{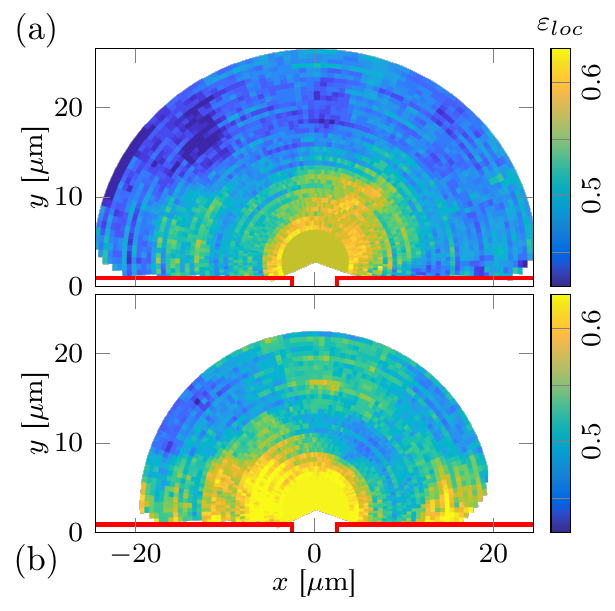}
\caption{Examples of 2D map of local porosity $\epsilon_{loc}$. (a) $I=0.5\,\text{mM}$ and $\Delta P=30\,\text{mbar}$; (b) $I=5\,\text{mM}$ and $\Delta P=10\,\text{mbar}$. Red lines represent pore edges. At the pore entrance, uniform patch is a consequence of the local porosity processing. The radial approach does not permit to detail the porosity at this location, and the porosity at $r=0$ is replicated on a surface corresponding to the surface of $60\,\mu$m$^2$ used to compute local porosity.}
\label{fig:porosity2D}
\end{figure}

The averaged porosity measurements presented above are not sufficient to understand in detail the clog microstructure and underlying physical and physico-chemical mechanisms at play in colloid assembly under filtration. We propose a systematic study of the local porosity $\varepsilon_{loc}$ of each clog. We define the local porosity at a given point  of the clog (in polar coordinates $\overrightarrow{r}$) as the porosity in a radially-oriented curved-trapezoidal box of approximately $60\,\mu$m$^2$ centered on $\overrightarrow{r}$ (corresponding to a 8-colloid wide box). Figure \ref{fig:porosity2D}\,(a) and \ref{fig:porosity2D}\,(b) show two examples for $(I=0.5\,\text{mM},\Delta P=30\,\text{mbar})$ and $(I=5\,\text{mM},\Delta P=10\,\text{mbar})$ respectively. One can observe qualitatively that the porosity decreases with the distance from the pore entrance, with a porosity divided by approximately 2. {This can be counter-intuitive as a compression of the clog could happen during the clogging process. We did not observe such a compression. It can be easily attributed to van der Waals adhesion of the particles which prevents them from moving.}


We perform an angular average at fixed $r$ of local porosity to obtain the porosity as a function of the distance $r$ to the pore entrance. Note that only locations whose absolute value of the $x$-projection of $\overrightarrow{r}$ is lower than half of inter-pore distance (25\,$\mu$m) are selected. This prevents taking into account parts of the cake in the ``influence region'' of a neighbour clog. {We made this first-order choice because we do not have information of the influence of clog interaction on their microstructure. Interactions between clogs (overlapping for instance) could locally affect the velocity field and so the local hydrodynamic forcing. This could lead to more dispersion of the results.} Figure \ref{fig:porosity_average} shows this quantity denoted $\langle\varepsilon_{loc}\rangle_\theta$ as a function of $r$ for $I=0.5\,$mM and $I=5\,$mM, and various $\Delta P$. When possible, this quantity is averaged over several clogs in the same experiment. One observes systematically a decrease of clog local porosity in the range $r\in[0,12]\,\mu$m.  Furthermore, figure \ref{fig:porosity_average}\,(right) does not  show significant difference when changing the applied pressure drop, especially at low $r$. For $I=0.5\,$mM, there is more variability of the local porosity for $r\in[0,12]\,\mu$m. It reveals the stochastic facet of colloid-surface adhesion and the initial steps of the clogging process, as discussed in previous works\cite{laar2016,dersoir2017,cejas2017}. The way and position first particles stick on the surface will affect the first portions of the clog and change locally the porosity (and so the global clog permeability). In summary, porosity is shown to be higher close to the pores, but without any strong systematic influence of ionic strength or applied pressure drop.

\subsubsection{Colloid spatial organisation}

\begin{figure}[h!]
\centering
\includegraphics[width=0.9\linewidth]{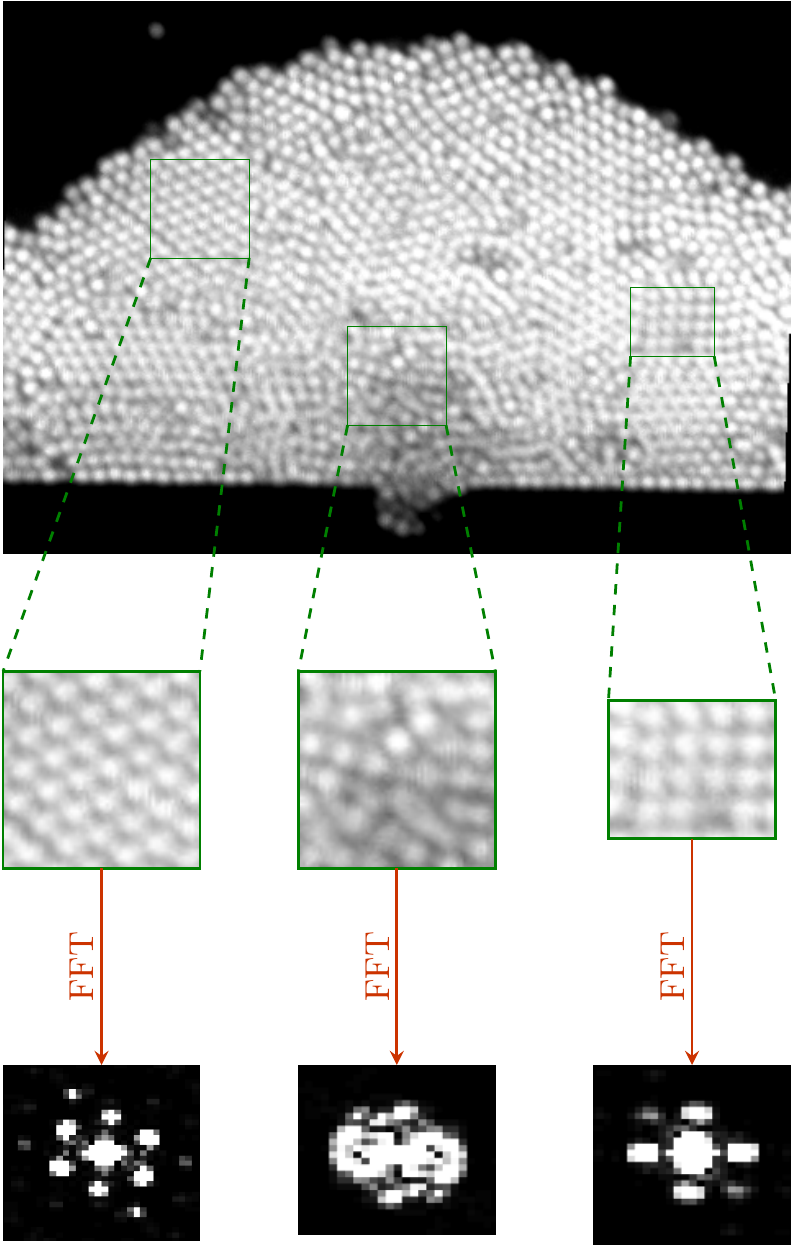}
\caption{Examples of 2D Fast Fourier Transform applied to different regions of a clog ($I=0.5\,$mM, $\Delta P=30$\,mbar). Predominately hexagonal, amorphous and square organization is seen, from left to right.}
\label{fig:TF}
\end{figure}

A first way to study local organisation of clogs consists in computing the 2D Fourier transform of different portions of the clogs' wall layer. Figure \ref{fig:TF} shows three examples of power spectra computed by 2D Fast Fourier Transform (FFT) on three different regions of a clog. One can differentiate three different local organisations: hexagonal, square and random. For each, the 2D FFT reveals a typical signature. Around the central spot, secondary spots separated by an angle specific to the lattice (60\textsuperscript{o} for hexagonal arrangement and 90\textsuperscript{o} for square arrangement) appear for the organized regions. On the contrary the more amorphous region reveals a kind of annular 2D FFT without spatial regularity.

The power spectrum obtained by 2D FFT will depend on the window dimension chosen for computation. To generalize this reciprocal-space analysis, we use a wavelet decomposition which is better suited for the local analysis aimed here. While Fourier decomposition uses a base with infinite-space support (sine and cosine), wavelet decomposition is a projection on a base composed of finite-space support elementary functions. Since we have a heterogeneous spatial organisation of the clogs, a Fourier transform does not allow a more quantitative local analysis of the local microstructure than figure \ref{fig:TF}. Continuous Wavelet Transform (CWT) overcomes this difficulty. Each point of the image and its surrounding environment can be decomposed in functions which generate a basis: the wavelets\footnote{They correspond to the sine and cosine functions in Fourier decomposition.}. They are localized both in space and frequency\cite{goupillaud1984}. Wavelet transformation is a widely used technique in many fields such as image processing\cite{manjunath1996} or glass structure analysis\cite{ding1998,harrop2002}. Each wavelet is built from a single ``mother'' function $\psi(\overrightarrow{X})$. A specific rotation, translation and expansion generates each wavelet for a given position $\overrightarrow{X}=(x,y)$ : 

\begin{equation}
    \Psi_{s,\Theta}(\overrightarrow{X})=\dfrac{1}{s}\psi\left(R_\Theta^{-1}\left(\dfrac{\overrightarrow{X}}{s}\right)\right),
\end{equation}

\noindent with

\begin{equation}
    R_\Theta=\begin{pmatrix}
    \cos\Theta & -\sin\Theta\\
    \sin\Theta & \cos\Theta
    \end{pmatrix}.
\end{equation}

\begin{figure}[h!]
\centering
\includegraphics[width=0.8\linewidth]{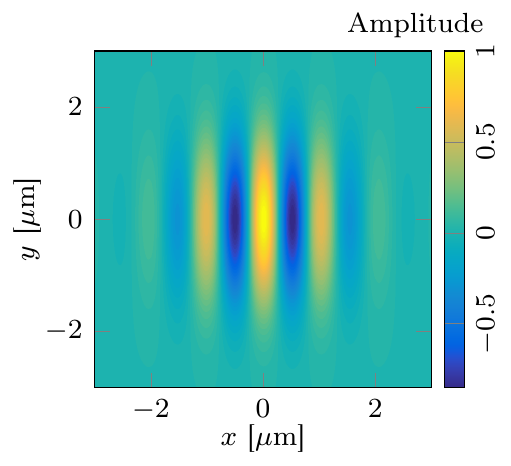}
\caption{Example of 2D Morlet wavelet used for wavelet analysis; $\Theta=0$ and $s=1\,\mu$m.}
\label{fig:morlet}
\end{figure}

\noindent $s$ is the expansion factor (or period) of the wavelet and $\Theta$ the rotation factor. We have chosen the classical 2D Morlet wavelet as the ``mother'' wavelet. It is a complex function which conserves phase information with good angular selectivity\cite{wang2009}. So it is a good candidate to study crystalline arrangements. A detailed explanation of the use of 2D-CWT is available in Chen \& Chu (2017)\cite{chen2017}. Figure \ref{fig:morlet} shows an example of 2D Morlet ``mother'' wavelet. 

\begin{figure}[h!]
\centering
\includegraphics[width=0.9\linewidth]{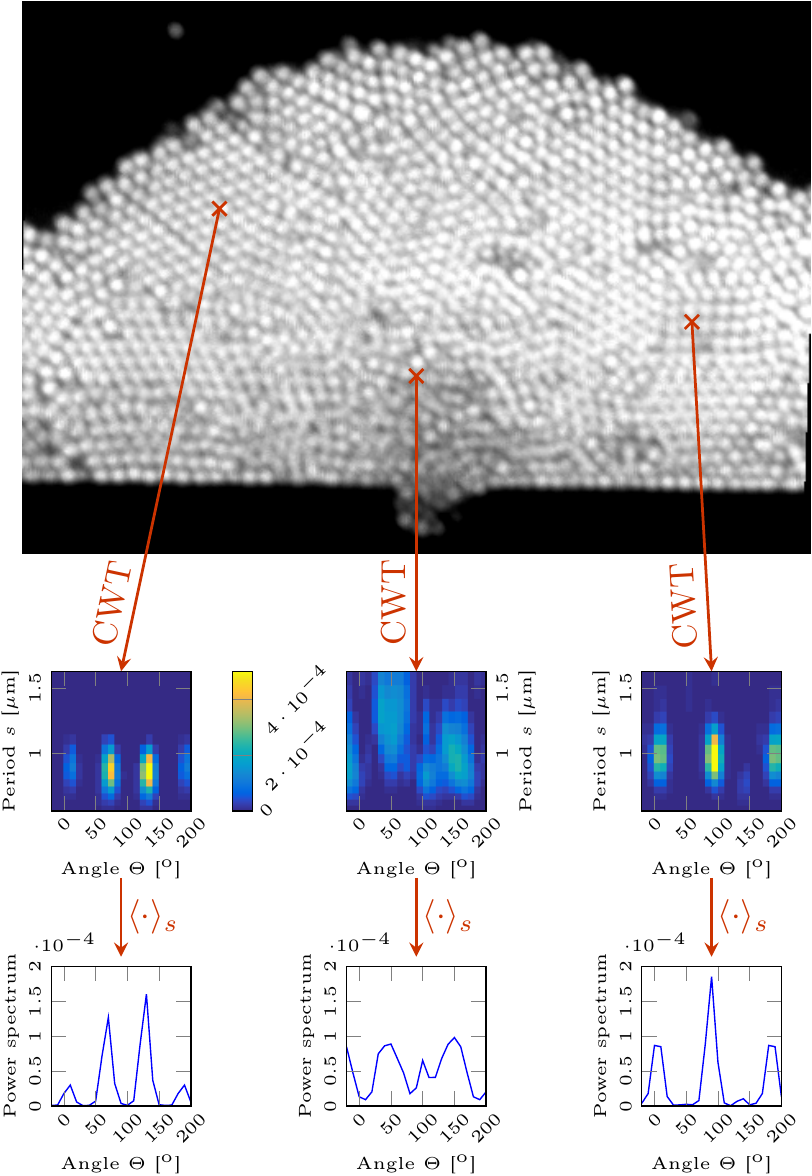}
\caption{Examples of Continuous Wavelet Transform applied to different locations of a clog ($I=0.5\,$mM, $\Delta P=30$\,mbar). The colorbar represents the power and is common to the three plots. $\langle\cdot\rangle_s$ means the average of the power spectrum amplitude on the period $s$.}
\label{fig:CWT}
\end{figure}

The result of applying a CWT on an image is a 2D map with angle $\Theta$ in abscissa and period $s$ in ordinate. Some examples are shown in figure \ref{fig:CWT}. Whereas for a Fourier decomposition, the power spectrum obtained from the transform can be plotted on a classic graph, coefficients derived from CWT are two-variable functions. Consequently, color on insets of figure \ref{fig:CWT} represents the power that is carried by each elementary function $\Psi_{s,\Theta}(\overrightarrow{X})$ for a given $(\Theta,s)$. The higher the power, the more important the corresponding elementary function contributes to the picture (2D signal). One can observe that for a point selected in an apparently hexagonal-lattice region, yellow spots appear with a periodicity of about 60\textsuperscript{o}. The period corresponds to the typical inter-particle distance, which is a bit under 1$\,\mu$m (one particle diameter). The same observation can be made for an apparently square-lattice region (periodicity of 90\textsuperscript{o}). For an amorphous region, no regular pattern appears on the CWT. Continuous Wavelet Transform thus enables extraction of quantitative information on local typical period and angular distributions.

\begin{figure}[h!]
\centering
\includegraphics[width=0.9\linewidth]{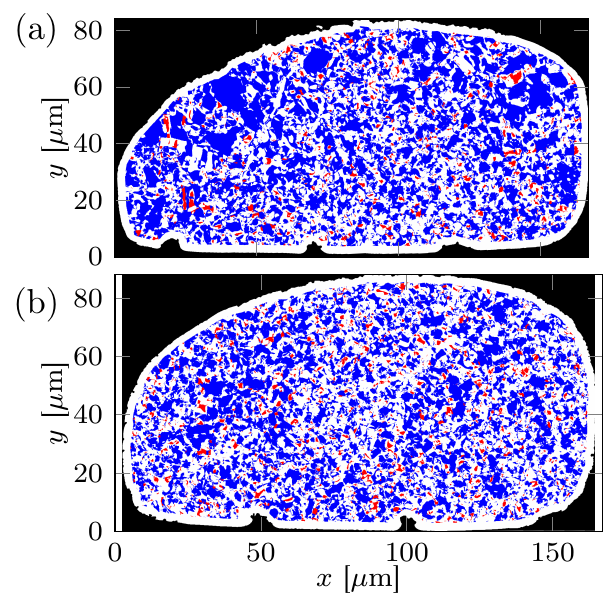}
\caption{Map of structures determined by CWT for $\Delta P=80\,$mbar, (a) $I=0.5\,$mM and (b) $I=5\,$mM. Blue regions correspond to hexagonal lattice and red ones to square lattice. White regions are amorphous. Black overhangs at the bottom of the cakes face pore entrances. Note that the white strip around the cake corresponds to a non-analyzed zone due to CWT edge effects.}
\label{fig:BBR}
\end{figure}

The results obtained from CWT can be averaged along ``period'' or ``angle'' direction. The ``period'' averaging gives a peak whose full width at half maximum gives some information on typical inter-particle distance distribution. That gives information very similar (not shown here) as the local porosity analysis (figure \ref{fig:porosity_average}). The average on the angle provides a succession of peaks, especially when a crystalline lattice is detected. Then we are able to compute the mean inter-peak distance. This quantity allows separation of three distinct regions: amorphous, hexagonal-lattice and square-lattice. Practically, hexagonal lattice corresponds to mean inter-peak angular distance in the range 55-65\textsuperscript{o} and square lattice corresponds to mean inter-peak angular distance in the range 85-95\textsuperscript{o}. Other angular distances are considered as related to amorphous regions. Figure \ref{fig:BBR} shows an example of a map obtained from this analysis (for two different ionic strengths). One can observe that hexagonal-lattice regions are largely predominant compared to square-lattice ones. Moreover, for $I=0.5\,$mM the hexagonal-lattice regions seem larger when the distance from the pores increases, whereas for $I=5\,$mM, they are quite homogeneously distributed.

In fact, maps presented in figure \ref{fig:BBR} are noisy. A lot of locations fit with crystalline zones whereas their typical size is not compatible with such a description. To denoise the maps, we apply an erosion-dilatation algorithm. Each blue or red connected region is eroded from the edge over 0.5$\,\mu$m (corresponding to one particle diameter total erosion). Remaining regions are then dilated by the same length. This allows us to isolate the sufficiently-extended crystalline-lattice regions. 

\begin{figure*}[h!]
\centering
\includegraphics[width=0.9\linewidth]{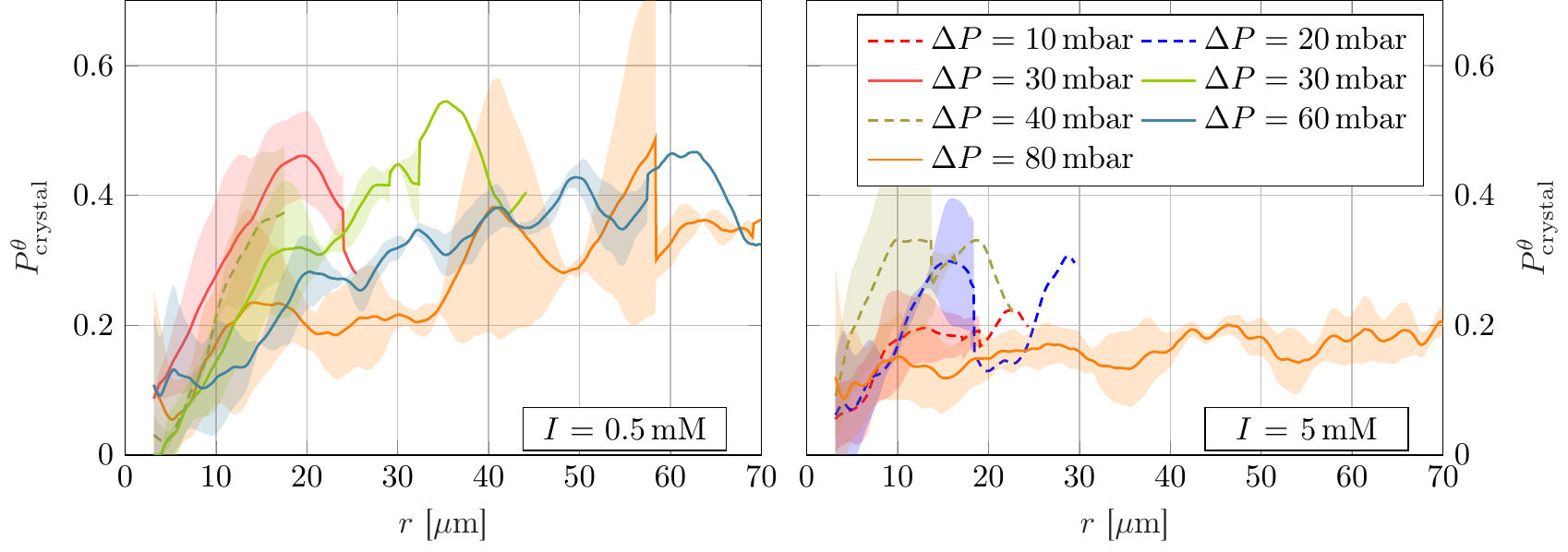}
\caption{Probability to encounter a crystalline region for (left) $I=0.5\,$mM and (right) $I=5\,$mM as a function of the distance from the pore entrance. The uncertainty range corresponds to standard deviation obtained with several clogs in the same experiment (when possible). Legend is common to both plots.}
\label{fig:crystalline_average}
\end{figure*}

To perform a local analysis of colloid arrangement in the filter cake, we adopt the same method as we used for porosity. We use polar coordinates with origin at pore entrance. We define shells of radius $r$ for $\theta\in[0,\pi]$ with thickness $\delta r\sim 6\mu$m. In this shell, we can define the probability to have points included in a crystalline region $P^\theta_{\text{crystal}}(r)$. Note that only locations whose absolute value of the $x$-projection of $\overrightarrow{r}$ is lower than half of inter-pore distance (25\,$\mu$m) are selected. Again, the basis for this is that it prevents taking into account parts of the cake in the ``influence region'' of a neighbour clog, to count one same position for two different clogs. Figure \ref{fig:crystalline_average} shows this quantity as a function of $r$ for (left) $I=0.5\,$mM and (right) $I=5\,$mM, and different $\Delta P$.

One observes again a systematic inhomogeneity of the clog microstructure. The probability to encounter crystalline regions increases with $r$ then reaches a plateau. As we move away from the pore entrance, the clog microstructure is more and more crystalline. For $I=0.5\,$mM, this plateau is between 0.3 and 0.4, and seems to decrease as $\Delta P$ rises. For $I=5\,$mM, the difference when changing $\Delta P$ is less clear, but globally the plateaus are around 0.2, lower than for $I=0.5\,$mM.

\section{Discussion: regimes of particle adhesion\label{sec:discussion}}

The results presented above are split in two categories: global approach and local analysis. The mean cake porosity is in the range 0.42--0.52 and does not depend on applied pressure drop or ionic strength. This averaged approach suggests that there is no influence of the forcing and physico-chemical properties on clog microstructure. However, the local analysis gives some decisive insights. Clogs are more porous at the vicinity of the pore entrance, then become more and more compact before reaching a plateau at a distance larger than $\sim12\,\mu$m from pore entrance. The porosity for small $r$ is clearly higher for low ionic strength than for $I=5\,$mM. This heterogeneity is totally concealed when one makes an average on the whole cake{'s wall layer}. 

Concerning the spatial organisation, a more refined study of radial distribution function shows a subtle effect: the screening length -- corresponding to the medium-range order -- is higher at low ionic strength and seems to decrease at high applied pressure drop. A local wavelet decomposition reveals crystalline regions -- mainly hexagonal lattices. The crystalline-region proportion increases with the distance from the pore and the plateau is higher for $I=0.5\,mM$. {Similar results were obtained numerically by Agbangla \emph{et al.}\cite{agbangla2014}. In the absence of inter-particle repulsive forces, particles are aggregated cake-like, without visible crystalline organisation. Addition of repulsive forces lead to formation of more organised arches of particles at the pore entrance.}

We propose an interpretation based on competition between hydrodynamic forcing and colloid-colloid repulsive interactions. To cross the energy barrier and cause adhesion of two colloidal particles, an external forcing is necessary. It can be provided by drag force due to hydrodynamic advection. {Note that this drag force should be influenced by hydrodynamic interactions\cite{dufresne2000,bhattacharya2005} which appear when a particle is travelling close to the wall or near another particle. It could affect the energy provided by the advection, at a distance up to several particle radii.} Extra energy can also come from Brownian motion. Let us consider a particle arriving in contact with the clog. We propose three different mechanisms: 

\begin{enumerate}
    \item \textit{ballistic regime}: the drag force applied on the impacting colloid is large enough to drive direct adhesion with colloid(s) in the clog;
    \item \textit{diffusive regime with adhesion}: \textcolor{black}{the drag force applied on the impacted colloid has decreased. Brownian motion however causes the particle to explore local configurations where the flow is still strong enough to lead to adhesion;}
    \item \textit{diffusive regime without adhesion}: there is not enough energy to observe colloid-colloid adhesion, the impacting colloid is simply constrained to be part of a ``repulsive glass''.
\end{enumerate}

The consequences on porosity and spatial organisation can be summed up as follows for each regime: 

\begin{enumerate}
    \item there is no time for particles to self-organize so porosity is high and structure is amorphous;
    \item Brownian motion allows a particle to explore a larger energy landscape before adhesion, and to find a more constrained position. This leads to less porous and more crystalline structure;
    \item the structure is similar to regime 2, but porosity could be slightly lower.
\end{enumerate}

{Let us insist about the role of Brownian motion. \textcolor{black}{It cannot bring extra energy sufficient to cross DLVO energy barrier and lead to particle adhesion: the typical Brownian motion energy is $\sim k_bT$ whereas the energy barrier is few tens of $k_bT$\cite{butt2003}. Consequently, in regime 2, adhesion is still provoked by the drag force.} But Brownian motion \textcolor{black}{acts} as a repulsive force allowing fluid to lubricate the interactions. Numerical simulations showed that Brownian motion facilitates yield of a colloidal gel under shear stress\cite{landrum2016,johnson2018}. This is the way a Brownian particle \textcolor{black}{could} access a larger energy landscape and help the clog self-organization. \textcolor{black}{This statement is rather difficult to access experimentally by the methods we used, and it should deserve a specific study with higher acquisition framerate, possibly higher spatial resolution and specific designs.}}

{Two other mechanisms could stimulate appearance and stability of the crystal zones, especially for the regime 3 (\textit{diffusive regime without adhesion}). Larsen \& Grier\cite{larsen1997} showed that long-range attractive interactions between like-charge colloids can allow the formation of metastable colloidal crystals. Moreover, presence of a wall lets appear attractive interactions between two colloids. Their study was made with very similar colloids (size, surface charge) as we used, but with an ionic strength 100 times lower than the ones we fixed (much larger repulsive interactions). Consequently we can assess that these mechanisms should play a role in the appearance of this third regime, when the relative contribution of repulsive interactions becomes dramatic.}

\begin{figure}[h!]
\centering
\includegraphics[width=1\linewidth]{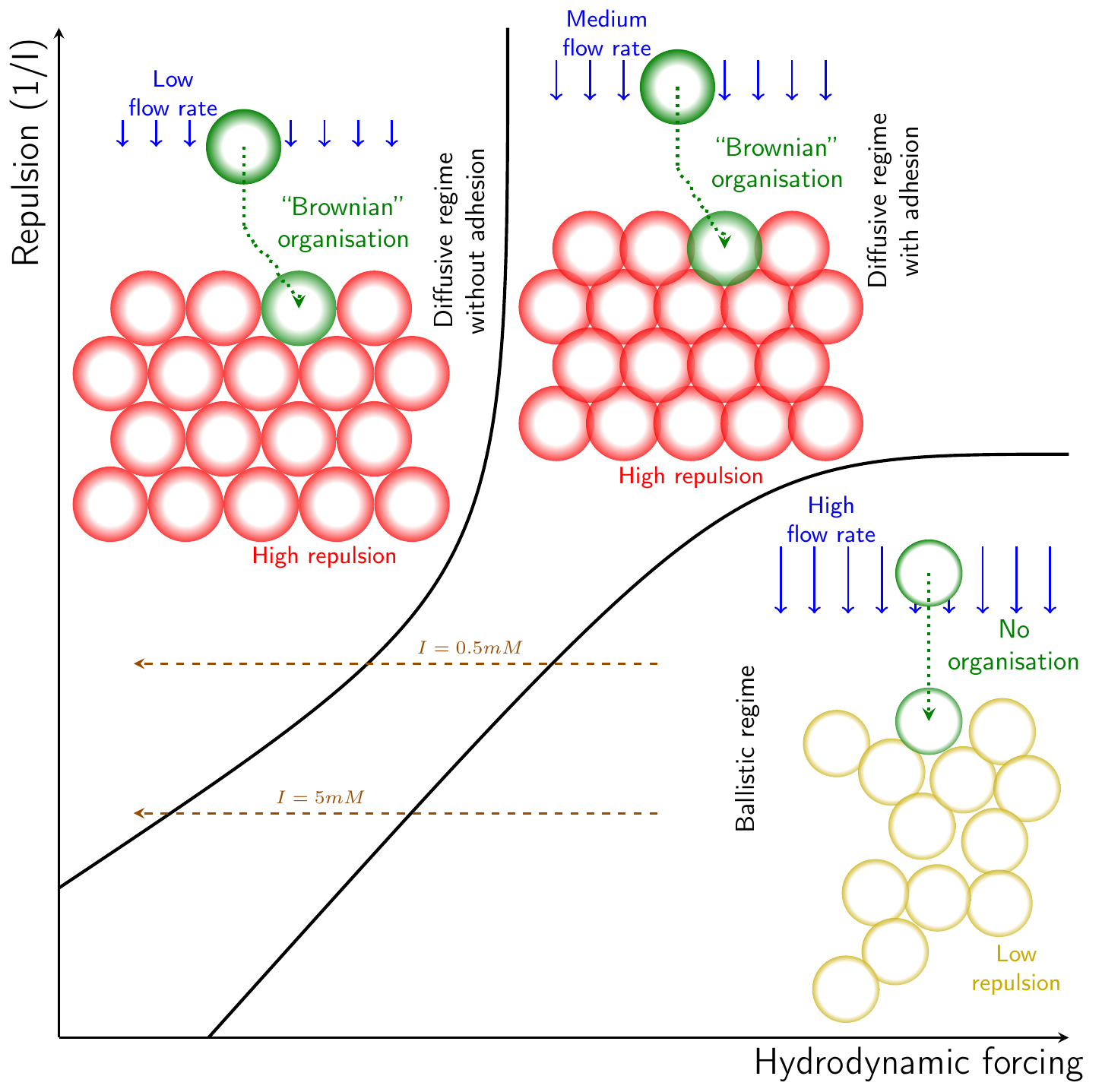}
\caption{Schematic diagram phase representing the three regimes for different ionic strength and hydrodynamic forcing. Shaded particle edges represent the EDL. Their overlapping mimes particle adhesion. Arrows represent possible displacements in the phase diagram during an experiment.}
\label{fig:diagramme}
\end{figure}

Figure \ref{fig:diagramme} shows a schematic phase diagram for these three regimes. Two parameters define the three regions of the phase diagram, corresponding to the three regimes described below: (i) hydrodynamic forcing (flow rate) and (ii) electric repulsive interactions (decreasing with increasing ionic strength). For weak repulsion and high hydrodynamic forcing, ballistic regime is dominant. For high repulsion and weak hydrodynamic forcing, diffusive regime without adhesion is dominant. For intermediate repulsion and hydrodynamic forcing, one can observe diffusive regime with adhesion. A change in Brownian motion intensity (determined by particle size, temperature, fluid viscosity) could slightly move the boundaries of the diagram.

We transpose this interpretation to our data. If we look at figure \ref{fig:RDF_fit}, the higher screening length (and hence the increased order) for $I=0.5\,$mM than for $I=5\,$mM could indicate a regime difference: diffusive (with or without adhesion) for $I=0.5\,$mM; ballistic for $I=5\,$mM. Furthermore, the plateau of $P^\theta_{\text{crystal}}(r)$ (figure \ref{fig:crystalline_average}) is higher for $I=0.5\,$mM than for $I=5\,$mM. This is consistent with a more organized microstructure, and so with the predominance of a diffusive regime at high interparticle electrostatic repulsion, and of the ballistic regime at high ionic strength. For $I=0.5\,mM$, the plateau seems lower as $\Delta P$ rises. This is also consistent with the appearance of a ballistic regime. 

We could hope for sharp transitions when changing ionic strength or pressure drop. Actually, hydrodynamic forcing conditions are time-dependent. Since we work at fixed pressure drop, the flow rate decreases as the clog grows. {Such situation where pressure drop is fixed can be encountered in real crossflow filtration devices.} Consequently, velocity (and so drag force) of an arriving particle will decrease with time. For a given experiment,
salt concentration is fixed, and thus the resulting double layer repulsion is also fixed. The phase diagram is traversed along a horizontal line from right to left. Two possible trajectories in the phase diagram are proposed in figure \ref{fig:diagramme} for two different ionic strengths. Consequently, a regime transition is expected during the clogging process. This is consistent with local analyses. The decrease of $\langle\varepsilon_{loc}\rangle_\theta$ with $r$ for both $I=0.5\,$mM and $I=5\,$mM is compatible with such a transition (see figure \ref{fig:porosity_average}). Moreover, the increase of $P^\theta_{\text{crystal}}$ with $r$, as shown in figure \ref{fig:crystalline_average}, reveals an increase of crystal regions proportion. This is also consistent with a transition inside the clog, from ballistic to one of the two diffusive regimes.

We globally observe some dispersion when changing the applied pressure drop. We should be able to extract a typical length of the local porosity decrease with $r$ -- and of the crystal proportion increase. Such a characteristic length should be dependent on $I$ and $\delta P$, which is not obvious on presented plots (figures \ref{fig:porosity_average} and \ref{fig:crystalline_average}). This assumption would be true if the first steps of the clogging process were reproducible among two experiments. Unfortunately, the stochastic facet of clogging makes a more accurate analysis difficult. The hydraulic resistance increase (and subsequent flow rate decline) can vary significantly from one experiment to another, depending on the local structure of the initial clog, next to the pore, which strongly influences the clog's hydrodynamic resistance.

\begin{figure}[h!]
\centering
\includegraphics[width=0.7\linewidth]{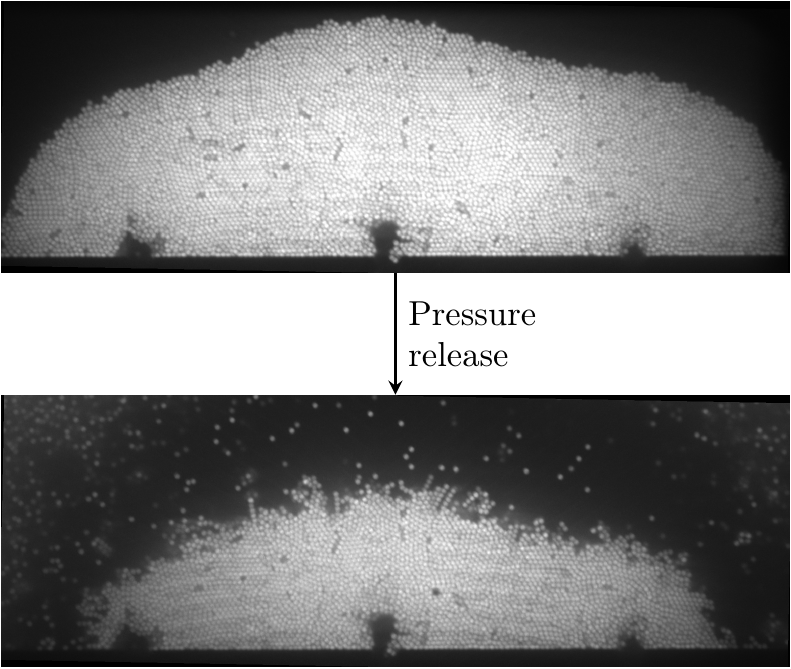}
\caption{Unclogging after pressure release. Clog before (top) and after (bottom) pressure release.}
\label{fig:unclogging}
\end{figure}

{Furthermore, regimes 2 and 3 are actually two sub-regimes of a more general \textit{diffusive regime}. We propose these two sub-regimes because of observations made after pressure release.} A release of hydrodynamic forcing revealed that a fraction of the clog remains stuck to the membrane (and therefore adhesive) whereas another part is re-suspended in the surrounding fluid, as presented in figure \ref{fig:unclogging}. {Particles still stuck after pressure release and, in addition to the amorphous region, large parts of the remaining clogs/cakes' wall layer have a crystalline microstructure -- revealing that in the crystalline zone, particles can be adhered or not.}

{Nevertheless}, we are not able to distinguish from the clog structure analysis the two diffusive regimes (with and without adhesion). Further experiments{, based on the \textit{unclogging} process,} are necessary to discriminate these regimes. This could be a way to refine our interpretation. {In addition, the fact that particles remain motionless after releasing the pressure drop is a clue that they are stuck to the glass plate.}

{An important point of our experiments concerns the absence of particle motion inside the clogs. As shown in figure \ref{fig:image_clog}, the particles belonging to a clog at a given time are at the same location at a later time. We never observed particle motion during the different image sequences acquired for this work. Such a motionless behaviour is also a clue of particle adhesion, or their belonging to a ``repulsive glass'' where hydrodynamic forcing freezes non-adhering particles in a given position -- the one they took up when they arrived at the clog. Moreover, this immobility is an indication that the fluid flowing through the clog does not affect clog microstructure over time. Particles belonging to an amorphous zone are indeed stuck to the surface and cannot move anymore.}

Note that these analyses are made using only the wall colloid layer. Although the glass plate at contact with the colloids could affect the absolute structure and organisation of this layer, the transitions and differences observed are consistent with a physical interpretation in the 3D bulk of the clog, and are reproducible. Moreover, if the glass plate was the main ingredient at play in wall-layer colloid organisation, it should make uniform the wall-layer microstructure. Unfortunately, the polystyrene particles used in these experiments are opaque and do not permit a 3D analysis using confocal microscopy. Working with optical-index-matched particles, which could enable analysis in the depth of clogs, is not at all straighforward without changing the interactions. As a matter of fact, van der Waals interactions, at play in the force balance (see DLVO in first section) are influenced by electric polarizability, related to the optical properties. Furthermore, real filtration devices can comprise lateral walls. This work proposes some insights about clog microstructure, and call other experimental studies to investigate the 3D structure {and understand the way particles pile up on the wall layer and affect its microstructure} -- use of opaque colloids did not allow a 3D study in our case.


\section{Conclusion and perspectives}
The filtration experiments we performed using microfluidic devices allow us to study the wall layer microstructure of clogs for two different parameters: applied pressure drop and ionic strength. We observed a clog development at the entrance of the pore, followed by growth of a filtration cake. Resolution was good enough to detect particles' positions. An analysis of clog porosity did not reveal strong global difference when changing pressure drop or salt concentration. At first glance, a study of colloid spatial organisation using radial distribution function also did not reveal clear influence of pressure drop and ionic strength. Nevertheless, by fitting these functions, we were able to extract a screening length which represents the medium-range order. This screening length is higher for low ionic strength, revealing a better spatial organisation.

A local analysis revealed spatial heterogeneity of the clogs. Porosity is higher at the vicinity of the pore then decreases with the distance from the pore. This observation is systematic for all pressure drops and ionic strengths. This heterogeneity is also valid for colloid spatial organisation. Fourier transform analysis of images of the clogs showed the presence of amorphous, square-lattice and hexagonal-lattice regions. Using 2D continuous wavelet transform, we were able to perform a local and systematic analysis of spatial organisation of the clogs. It revealed a large predominance of hexagonal-lattice regions compared to square-lattice ones. Moreover, the proportion of crystalline regions increases with the distance from the pore entrance before reaching a plateau. It is higher for high interparticle repulsion, and seems to be lower as pressure drop rises. 

We gathered all these observations in a new framework based on a phase diagram. Three regimes are accessible, depending on flow and repulsive interaction intensity. One is a ballistic regime, where addition of a colloid to a clog is the result of a direct adhesion due to drag force. Two diffusive regimes (with or without adhesion) are due to lower hydrodynamic forcing or higher repulsion. Whereas the ballistic regime does not allow organisation of the particles, the two others let allow time to an arriving colloid to ``plug the holes''. \textcolor{black}{Where there is adhesion, it is always due to drag forces. Brownian motion can only help to organize the clog.} This framework is compatible with all our experimental observations. {These results could be compared with the distinction between diffusion-limited aggregation (DLA) and reaction-limited aggregation (RLA)\cite{lin1989}. DLA is a rapid  
process in which particles immediately adhere to the aggregate, and provides loosely packed aggregates, whereas RLA is a slower process that requires each particle to make multiple attempts, allowing exploration of the existing aggregate and results in a denser (higher fractal dimension) aggregates.}


{Several} kinds of experiments could complete this study: 3D analysis using confocal microscopy (but that would require index matching, which is not straightforward), X-ray microtomography {and different chip designs to observe clogs from the side}, in order to refine our results in the third dimension; unclogging analysis to distinguish the two diffusive regimes; {acquisitions with higher framerate and spatial resolution to observe the subtle effect of Brownian motion on clog self-organization}. The second point could be the opportunity to study the unclogging dynamics when hydrodynamic forcing is released.

\section*{Conflicts of interest}
There are no conflicts to declare.

\section*{Acknowledgements}
We acknowledge the F\'ed\'eration FERMAT and University of Toulouse (Project NEMESIS) for funding these researches. This work was partly supported by LAAS-CNRS micro and nanotechnologies platform member of the French RENATECH network. We acknowledge Julien Lefort and Maëlle Ogier for technical and microscopy support. We warmly thank Patrice Bacchin and Paul Duru for the fruitful discussions and their manuscript revise.



\balance


\bibliography{Clogging} 
\bibliographystyle{rsc} 

\end{document}